\documentclass[twocolumn,english,prl]{revtex4}
\usepackage[T1]{fontenc}
\usepackage[latin9]{inputenc}
\usepackage{babel}
\usepackage{float}
\usepackage{mathrsfs}
\usepackage{amsmath}
\usepackage{amssymb}
\usepackage{graphicx}
\usepackage{esint}
\usepackage[unicode=true,pdfusetitle,
 bookmarks=true,bookmarksnumbered=false,bookmarksopen=false,
 breaklinks=false,pdfborder={0 0 1},backref=false,colorlinks=false]
 {hyperref}
\usepackage{breakurl}

\makeatletter
\@ifundefined{textcolor}{}
{%
 \definecolor{BLACK}{gray}{0}
 \definecolor{WHITE}{gray}{1}
 \definecolor{RED}{rgb}{1,0,0}
 \definecolor{GREEN}{rgb}{0,1,0}
 \definecolor{BLUE}{rgb}{0,0,1}
 \definecolor{CYAN}{cmyk}{1,0,0,0}
 \definecolor{MAGENTA}{cmyk}{0,1,0,0}
 \definecolor{YELLOW}{cmyk}{0,0,1,0}
}

\makeatother

\begin{document}

\title{Floquet-Bloch theory and topology in periodically driven lattices}

\author{A. Gómez-León}

\affiliation{Instituto de Ciencia de Materiales de Madrid (ICMM-CSIC), Cantoblanco,
28049 Madrid, Spain}

\author{G. Platero}

\affiliation{Instituto de Ciencia de Materiales de Madrid (ICMM-CSIC), Cantoblanco,
28049 Madrid, Spain}
\begin{abstract}
We propose a general framework to solve tight binding models in D
dimensional lattices driven by ac electric fields. Our method is valid
for arbitrary driving regimes and allows to obtain effective Hamiltonians
for different external fields configurations. We establish an equivalence
with time independent lattices in D+1 dimensions, and analyze their
topological properties. Further, we demonstrate that non-adiabaticity
drives a transition from topological invariants defined in D+1 to
D dimensions. Our approach provides a theoretical framework to analyze
ac driven systems, with potential applications in topological states
of matter, and non-adiabatic topological quantum computation, predicting
novel outcomes for future experiments.
\end{abstract}

\date{\today}

\maketitle

\paragraph{Introduction:}

Periodically driven quantum systems has been a fastly growing research
field in the last years. The development of effective Hamiltonians
describing ac driven systems at certain regimes, has allowed to predict
novel properties such as topological phases\cite{photinduced-top-transition,Floquet-top-ins,Cirac-Maj-Floq},
and quantum phase transitions\cite{Dicke-Model,Ising-Model} that
otherwise, would be impossible to achieve in the undriven case. Therefore,
the application of ac fields has become a very promising tool to engineer
quantum systems. On the other hand, the obtention of effective Hamiltonians
can be a difficult task, depending on the driving regime to be considered.

In this work, we provide a general framework to study periodically
driven quantum lattices. By means of our approach, it is possible
to solve with arbitrary accuracy their time evolution, and obtain
effective Hamiltonians for the different driving regimes. We consider
solutions of Floquet-Bloch form, based on the symmetries of the system,
and characterize the states in terms of the quasi-energies, which
are well defined for all driving regimes. The states belong to a composed
Hilbert space, in which time is treated as a parameter. As we will
see below, it allows to formally describe the ac driven D dimensional
lattice, as analog to a time independent D+1 dimensional lattice.
This description enlightens the underlaying structure of periodically
driven systems, in which the initial Bloch band splits into several
copies (Floquet-Bloch bands), where the coupling between them directly
depends on the driving regime. Interestingly, the isolated Floquet-Bloch
bands possess the same topological properties as the Bloch bands of
the undriven system, now tuned by the external field parameters. Thus,
the topological invariants for the isolated Floquet-Bloch bands can
be obtained following the general classification of time independent
systems (AZ classes\cite{Zirnbauer,topological-clasification,Altland}).
However, we also demonstrate that lowering the frequency, the bands
couple to each other. In that case, the topological structures are
classified according to a base manifold of dimension D+1. In consequence,
one can simulate higher dimensional tight binding (TB) models with
exotic tunable hoppings by just coupling the system to ac electric
fields.

Our approach is valid for arbitrary dimension, and it allows to independently
analyze the effect of the field amplitude and frequency. We show that
the field amplitude controls the renormalization of the system parameters,
while the frequency acts analogously to a DC electric field in the
extra dimension. In particular, this last property relates the high
frequency regime with the existence of Bloch oscillations and Landau-Zener
transitions between bands\cite{Zener}, establishing a direct relation
between diabatic regime and localization\cite{Localization-Berryphases}.
We illustrate our formalism with the analysis of an ac driven dimers
chain\cite{polyacetylene1,polyacetylene2,Pierre-Dimers}.
\begin{figure}[h]
\includegraphics[scale=0.35]{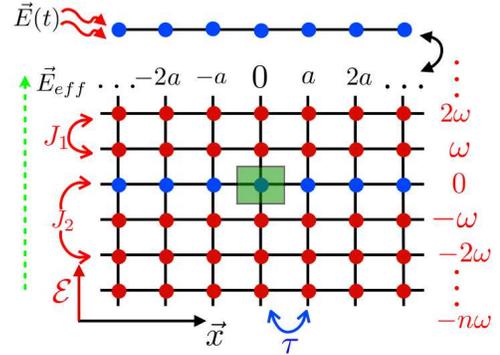}

\caption{\label{fig:Space-time lattice-1-1}Schematic figure of the equivalence
between a periodically driven 1D chain, and the effective time independent
lattice in 2D. We plot the positions of the undressed states (blue)
and their spatial distribution at sites $ra$, $r\in\mathbb{Z}$.
Each site is coupled to a set of dressed states (red color), with
coupling $\propto\tau J_{p}\left(A_{0}\right)$, being $p=n-m$ the
difference in the number of photons between the coupled Floquet bands,
and $A_{0}$ the vector potential amplitude. We draw the Wigner-Seitz
unit cell in 2D (green), and the effective dc electric field (see
text) along the energy axis $\mathcal{E}$ (green dotted arrow). The
extra dimension $\mathcal{E}$ arises due to the time periodicity.}
\end{figure}

\paragraph{Theory:}

We start by considering a Hamiltonian with lattice and time translation
invariance: $H\left(\mathbf{x}+\mathbf{a}_{i},t+T\right)=H\left(\mathbf{x}+\mathbf{a}_{i},t\right)=H\left(\mathbf{x},t+T\right)$,
characterized by lattice vectors $\mathbf{a}_{i}$ and time period
$T=2\pi/\omega$. Under these assumptions, we assume the Floquet-Bloch
ansatz\cite{Grifoni-Floquet,Gloria-Floquet} $|\Psi_{\alpha,\mathbf{k}}\left(\mathbf{x},t\right)\rangle=e^{i\mathbf{k}\cdot\mathbf{x}-i\epsilon_{\alpha,\mathbf{k}}t}|u_{\alpha,\mathbf{k}}\left(\mathbf{x},t\right)\rangle$,
being $\epsilon_{\alpha,\mathbf{k}}$ the quasi-energy for the $\alpha$
Floquet state, and $\mathbf{k}$ the wave-vector. The Floquet-Bloch
states $|u_{\alpha,\mathbf{k}}\left(\mathbf{x},t\right)\rangle$ are
periodic in both $\mathbf{x}$ and $t$, and belong to the composed
Hilbert space (Sambe space) $\mathcal{S}=\mathscr{H}\otimes\mathcal{T}$\cite{Sambe},
where $\mathcal{T}$ is the space of T-periodic functions and $\mathscr{H}$
the Hilbert space. It defines a composed scalar product $\langle\langle\ldots\rangle\rangle=\int_{0}^{T}\langle\ldots\rangle dt/T$.
In this space, the Floquet states fulfill the orthogonality condition
$\langle\langle u_{\alpha}\left(t\right)|u_{\beta}\left(t\right)\rangle\rangle=\frac{1}{T}\sum_{n,m}\int_{0}^{T}e^{i\omega\left(n-m\right)t}\langle u_{\alpha,n}|u_{\beta,m}\rangle dt=\delta_{\alpha,\beta}\delta_{n,m}$,
and the Floquet equation:
\begin{eqnarray}
\mathcal{H}\left(\mathbf{k},t\right)|u_{\alpha,\mathbf{k}}\rangle & = & \epsilon_{\alpha,\mathbf{k}}|u_{\alpha,\mathbf{k}}\rangle,\label{eq:Floquet-Eq}\\
\mathcal{H}\left(\mathbf{k},t\right) & \equiv & e^{-i\mathbf{k}\cdot\mathbf{x}}\left(H\left(t\right)-i\partial_{t}\right)e^{i\mathbf{k}\cdot\mathbf{x}}\nonumber \\
 & = & H_{\mathbf{k}}\left(t\right)-i\partial_{t}\label{eq:Floquet-operator}
\end{eqnarray}
where $|u_{\alpha,n}\rangle$ are the coefficients of the Fourier
expansion in $t$, and $\mathcal{H}\left(\mathbf{k},t\right)$ the
Floquet operator. We also define the Floquet-Bloch annihilation and
creation operators $\left\{ c\left(t\right)_{\alpha,\mathbf{k}},c\left(t\right)_{\alpha,\mathbf{k}}^{\dagger}\right\} $.
Note that the Floquet ansatz allows to map the time dependent Schrödinger
equation to an eigenvalue equation, in which $t$ and $\mathbf{k}$
are both parameters. We propose below that an exact mapping between
a D dimensional ac driven system and a D+1 undriven one can be established,
in which $\mathcal{H}\left(\mathbf{k},t\right)$ is equivalent to
an static Hamiltonian. It will allow to classify the topological invariants
in terms of the mappings from the parameter space to the set of Floquet
operators $\mathcal{H}\left(\mathbf{k},t\right)$, being the parameter
space now given by $S^{1}\times\mathbb{T}^{n}$, where $n$ is the
dimension of the First Brillouin zone (FBZ).

We first define the Fourier transforms:
\begin{eqnarray}
c\left(t\right)_{\alpha,\mathbf{k}} & = & N^{-D/2}\sum_{j=1}^{N}\sum_{n=-\infty}^{\infty}c{}_{\alpha,j,n}e^{i\mathbf{k}\cdot\mathbf{R}_{j}+in\omega t}\label{eq:Fourier-transf}\\
c\left(t\right)_{\alpha,\mathbf{k}}^{\dagger} & = & N^{-D/2}\sum_{j=1}^{N}\sum_{n=-\infty}^{\infty}c{}_{\alpha,j,n}^{\dagger}e^{-i\mathbf{k}\cdot\mathbf{R}_{j}-in\omega t},\nonumber 
\end{eqnarray}
where $N$ is the number of sites in the the lattice with periodic
boundary conditions and $D$ the dimension of the undriven system.
For a time dependent Hamiltonian in the dipolar approximation, the
relation between undriven and driven system is given by the minimal
coupling $\mathbf{k}\rightarrow\mathbf{K}\left(t\right)=\mathbf{k}+\mathbf{A}\left(t\right)$,
where $\mathbf{A}\left(t\right)$ is the vector potential. Equivalently,
one can notice that in the undriven TB Hamiltonian, the hopping parameters
$\tau_{j,l}$ are $k$ independent. Thus, one can obtain the time
dependent TB by using the minimal coupling in the inverse Fourier
transform, leading to the time dependent hoppings $\tau_{j,l}\left(t\right)=\tau_{j,l}e^{i\mathbf{A}\left(t\right)\cdot\left(\mathbf{R}_{j}-\mathbf{R}_{l}\right)}$,
and being the Hamiltonian

\begin{equation}
H_{\mathbf{k}}\left(t\right)=\sum_{\alpha,\mathbf{k}}\sum_{j,l}\tau_{j,l}\left(t\right)e^{i\mathbf{k}\cdot\left(\mathbf{R}_{j}-\mathbf{R}_{l}\right)}c\left(t\right)_{\alpha,\mathbf{k}}^{\dagger}c\left(t\right)_{\alpha,\mathbf{k}},\label{eq:H-minimal-coupling}
\end{equation}
where the $\alpha$ index labels the Floquet state, and the time dependence
in the operators is included. Including the expansion in Fourier series
of $c\left(t\right)_{\alpha,\mathbf{k}}$, and $c\left(t\right)_{\alpha,\mathbf{k}}^{\dagger}$
(Eq.\ref{eq:Fourier-transf})
\begin{equation}
H_{\mathbf{k}}\left(t\right)=\sum_{\alpha,\mathbf{k}}\sum_{n,m}\sum_{j,l}\tau_{j,l}\left(t\right)c_{\alpha,\mathbf{k},n}^{\dagger}c_{\alpha,\mathbf{k},m}e^{i\kappa\left(t\right)\cdot\left(\rho_{n,j}-\rho_{m,l}\right)},\label{eq:Floquet-tight-Fourier}
\end{equation}
being the quadrivectors $\kappa\left(t\right)\equiv\left(-t,\mathbf{k}\right)$
and $\rho_{n,j}\equiv\left(n\omega,\mathbf{R}_{j}\right)$. Eq.\ref{eq:Floquet-tight-Fourier}
gives a description of the time dependent Hamiltonian in terms of
the time independent operators $\left\{ c_{\alpha,\mathbf{k},n},c_{\alpha,\mathbf{k},n}^{\dagger}\right\} $.
Finally, the use of the composed scalar product allows to obtain the
quasi-energies by diagonalization of the matrix: 
\begin{gather}
\langle\langle u_{\alpha,\mathbf{k},n}|\mathcal{H}\left(\mathbf{k},t\right)|u_{\alpha,\mathbf{k},m}\rangle\rangle=\tilde{\tau}_{n,m}-n\omega\delta_{n,m},\label{eq:Floquet-tight-binding}\\
\tilde{\tau}_{n,m}\equiv\frac{1}{T}\int_{0}^{T}\sum_{j,l}^{N}\tau_{j,l}\left(t\right)e^{i\kappa\left(t\right)\cdot\left(\rho_{n,j}-\rho_{m,l}\right)}dt,\label{eq:hoppings-nm}
\end{gather}
where $n\omega\delta_{n,m}$ is the Fourier space representation of
$-i\partial_{t}$. Eq.\ref{eq:Floquet-tight-binding} is analog to
a time independent TB in D+1 dimensions with an electric field of
unit intensity applied along the extra dimension, and sites labeled
by $\left(n,j\right)$ (see Fig.\ref{fig:Space-time lattice-1-1}).
The effective electric field breaks translational symmetry in the
$\mathcal{E}$ axis, and one can differentiate two regimes: Low and
high frequency.

In the low frequency regime ($\omega\ll\tau_{j,i}$), it is a good
approximation to neglect the effect of the time derivative in the
Floquet operator (Eq.\ref{eq:Floquet-operator}), or equivalently,
the Stark shift due to the effective electric field. Then, we restore
the $\mathcal{E}$ axis translational symmetry, and the operator can
be diagonalized by Fourier transform of Eq.\ref{eq:Floquet-tight-binding}
to $t$ domain. The obtained $\mathcal{H}\left(\mathbf{k},t\right)$
is analog to a Hamiltonian over a D+1 compact base manifold, that
we define as the First Floquet Brillouin Zone (FFBZ), parametrized
by $\left\{ t,\mathbf{k}\right\} \in S^{1}\times\mathbb{T}^{n}$ \cite{Sup.inf.}.
Hence, the system topology is classified according to the AZ class
of D+1 static Hamiltonians. Note that if one initially assumes adiabatic
evolution, the Floquet structure, the parameters renormalization by
the field amplitude, and the additional dimension of the base manifold
are not obtained. Thus, the topological classification is not clearly
stablished.

For high frequency ($\omega\gg\tau_{j,l}$), the effective electric
field produces Bloch-Zener transitions and Bloch oscillations\cite{Zener,Localization-Berryphases},
inducing localization in $n\omega$, and decoupling the Floquet bands.
Then, Eq.\ref{eq:Floquet-tight-binding} becomes block-diagonal in
Fourier space, and the effective Floquet operator time independent.
In addition, it is defined over a D dimensional base manifold $\mathbf{k}\in\mathbb{T}^{n}$
(FBZ). In consequence, the topological classification is given by
the AZ classes for time independent systems in D dimensions. The transition
from a D+1 to a D dimensional base manifold (equivalently, from the
FFBZ to the FBZ), as one increases the frequency $\omega$, is driven
by the non-adiabatic processes along the $t$ axis\cite{Non-adiabatic}.

Importantly, note that in general the hopping between $n,m$ neighbors
depends on the amplitude of the vector potential (Eq.\ref{eq:hoppings-nm}).
It allows to design ``effective lattices'' by tuning the hoppings
with the ac field\cite{NNN}.

\paragraph{Periodically driven dimers chain:}

\begin{figure}[h]
\includegraphics[scale=0.25]{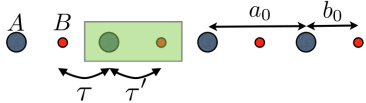}

\caption{\label{fig:Schematic-dimer1}Schematic figure for a dimers chain with
two inequivalent atoms (A, B) for unit cell (green area). $b_{0}$
is the distance between A-B atoms within the same primitive cell,
$a_{0}$ the lattice translation vector, $\tau^{\prime}$ the hopping
within the same dimer, and $\tau$ the hopping to the next one.}
\end{figure}

Here we consider the case of a dimers chain coupled to an ac electric
field, with hoppings $\tau$ and $\tau^{\prime}$, and periodic boundary
conditions (Fig.\ref{fig:Schematic-dimer1}).The electric field $E\left(t\right)=-\partial_{t}A\left(t\right)$,
is given by the vector potential $A\left(t\right)=A_{0}\sin\left(\omega t\right)$,
where $A_{0}\equiv qE_{0}/\omega$ (we fix $q=-1$).

By calculating $\tilde{\tau}_{n,m}$ from Eq.\ref{eq:hoppings-nm},
one obtains:
\begin{eqnarray}
\tilde{\tau}_{n,m} & = & \tau\left(\begin{array}{cc}
0 & \rho_{F}\left(k\right)\\
\tilde{\rho}_{F}\left(k\right) & 0
\end{array}\right),\label{eq:Dimers-FHamiltonian}\\
\rho_{F}\left(k\right) & \equiv & \lambda e^{-ikb_{0}}J_{n-m}\left(A_{0}b_{0}\right)\nonumber \\
 &  & +e^{ik\left(a_{0}-b_{0}\right)}J_{m-n}\left(A_{0}\left(a_{0}-b_{0}\right)\right),\nonumber \\
\tilde{\rho}_{F}\left(k\right) & \equiv & \lambda e^{ikb_{0}}J_{m-n}\left(A_{0}b_{0}\right)\nonumber \\
 &  & +e^{-ik\left(a_{0}-b_{0}\right)}J_{n-m}\left(A_{0}\left(a_{0}-b_{0}\right)\right).\nonumber 
\end{eqnarray}
where $\lambda=\tau^{\prime}/\tau$, and $\left(n,m\right)\in\mathbb{Z}$.
In contrast with the undriven case, the spectrum depends on the intra-dimer
distance $b_{0}$, and the hoppings are renormalized by the field
amplitude. Note that for the limit of vanishing ac-field $A_{0}\rightarrow0$,
the quasi-energies match the energies of the undriven system.

We first consider the high frequency regime ($\omega\gg\tau,\tau^{\prime}$),
and select the Floquet band $n=m=0$. In this case, the Floquet operator
is block diagonal, and the system can be described by a time independent
2 by 2 matrix 
\begin{equation}
\mathcal{H}_{k}^{\left(0\right)}=\tau\vec{g}\left(k\right)\cdot\vec{\sigma},\label{eq:Zeroth-order-H}
\end{equation}
where $\vec{g}\left(k\right)=\left(\Re\left(\tilde{\rho}_{F}\right),\Im\left(\tilde{\rho}_{F}\right),0\right)$
for $n=m=0$, and $\vec{\sigma}=\left(\sigma_{x},\sigma_{y},\sigma_{z}\right)$
are the Pauli matrices.

In order to compare with the case with periodic boundary conditions
discussed above, we also consider a finite dimers chain with Hamiltonian
$H\left(t\right)=H_{0}+qE\left(t\right)x$, being $H_{0}$ the time
independent TB Hamiltonian, and $qE\left(t\right)x$ the coupling
with the electric field (details in \cite{Sup.inf.}).

\begin{figure}[h]
\includegraphics[scale=0.6]{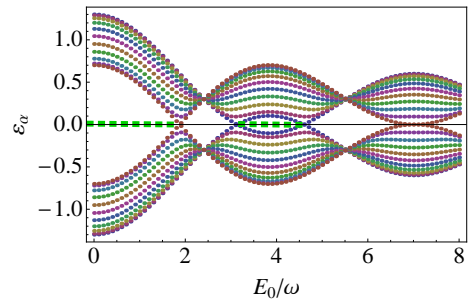}

\caption{\label{fig:Quasi-energies-vs-Intensity1}Quasi-energy spectrum vs
$E_{0}/\omega$ for $\lambda=0.3$, and $b_{0}=0$, considering Eq.\ref{eq:Zeroth-order-H}
(high frequency). The band structure is obtained by considering 10
$k$ vectors equally spaced within the FBZ. We also included in green
color the gapless modes obtained from the numerical calculation, $\omega=10$.
All parameters in units of $\tau$.}
\end{figure}

Fig.\ref{fig:Quasi-energies-vs-Intensity1} shows the quasi-energy
spectrum in high frequency regime. We included in green dotted lines
the regions of existence of edge states, obtained from the numerical
calculation of the finite size system.

\begin{figure}[h]
\includegraphics[scale=0.75]{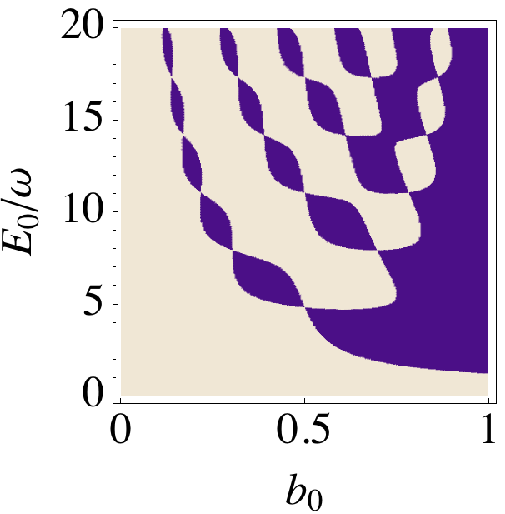}\includegraphics[scale=0.75]{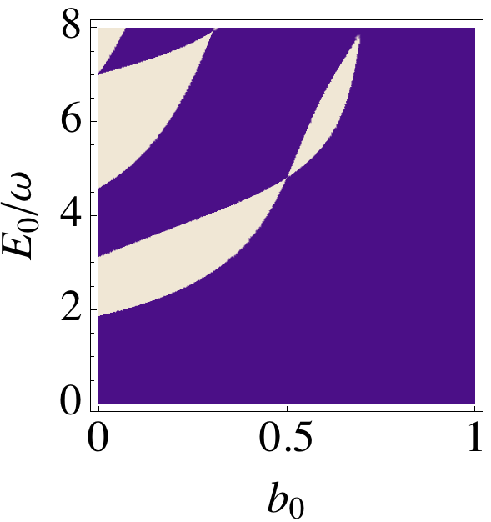}

\caption{\label{fig:Topological-phase-diagram}Topological phase diagram in
the high frequency regime for an ac driven dimers chain. We considered
$\lambda=1.5$ (left), and $\lambda=0.3$ (right). Dark color area
means $\nu_{1}=\pi$, and light area means $\nu_{1}=0$. Note that
even for $\lambda>1$ we can induce a non-trivial topology, in contrast
with the undriven case. Further, the phase diagram for $\lambda=0.3$
(right) agrees with the existence of edge states in Fig.\ref{fig:Quasi-energies-vs-Intensity1}
($b_{0}=0$).}
\end{figure}

The appearance of zero energy modes is a finite size effect linked
to the underlaying topology of the system. The relation between them
and the bulk topology is via the bulk to edge correspondence, which
relates the number of zero energy modes at the boundary, carrying
a topological number, with the bulk topological invariants\cite{Band-Inversion-TI}.
Therefore, the calculation of the topological invariants of $\mathcal{H}_{k}^{\left(0\right)}$
should predict their existence in this regime (Fig.\ref{fig:Quasi-energies-vs-Intensity1}).
$\mathcal{H}_{k}^{\left(0\right)}$ belongs to the BDI class (as the
one corresponding to the undriven system), with time reversal, particle-hole,
and chiral symmetry\cite{topological-clasification}. In 1D, the BDI
class is characterized by a winding number $\nu_{1}$, which classifies
mappings $\mathbb{T}^{1}\rightarrow\mathbb{R}^{2}-\left\{ 0\right\} \simeq S^{1}$,
from the FBZ to the family of Hamiltonians $\mathcal{H}_{k}^{\left(0\right)}$:

\begin{eqnarray}
\nu_{1} & = & \oint\langle u_{\alpha,k}|i\partial_{k}|u_{\alpha,k}\rangle dk\nonumber \\
 & = & \frac{\pi}{2}\left(1+\text{sign}\left(J_{0}^{2}\left(y\right)-\lambda^{2}J_{0}^{2}\left(x\right)\right)\right),\label{eq:Phase-diagram-dimers}
\end{eqnarray}
where $y\equiv A_{0}\left(a_{0}-b_{0}\right)$, $x\equiv A_{0}b_{0}$,
and $|u_{\alpha,k}\rangle$ are the closed lifts of $\mathcal{H}_{k}^{\left(0\right)}$.
Eq.\ref{eq:Phase-diagram-dimers} shows, that in contrast with the
undriven case\cite{Pierre-Dimers}, one can create non-trivial topological
phases even for $\lambda>1$, where the undriven system is in the
trivial phase (Fig.\ref{fig:Topological-phase-diagram} left). This
is an example of topology induced by the driving.

In Fig.\ref{fig:Topological-phase-diagram} (right) we plot the phase
diagram corresponding to $\lambda=0.3$, which correctly predicts
the existence of edge states for $b_{0}=0$ (Fig.\ref{fig:Quasi-energies-vs-Intensity1}).

In summary, we have shown that in the high frequency regime, the topological
properties can be obtained using an effective static Hamiltonian $\mathcal{H}_{k}^{\left(0\right)}$,
and that they can be tuned by the field amplitude.

As we decrease $\omega$, the different Floquet bands couple to each
other, and the isolated band picture is not accurate. In this regime,
one must consider the full Floquet operator (Eq.\ref{eq:Floquet-Eq}),
which for this system is not exactly solvable. Due to the coupling
between Floquet bands, two different but related effects happen as
$\omega$ is reduced: Bands inversions, and the emergence of a D+1
parameter space. 

Bands inversions correspond to crossings of the bands, in which the
symmetry is exchanged (e.g., it occurs in quantum wells of HgTe/CdTe
as the well thickness reaches a critical value\cite{Band-Inversion-TI}).
This effect happens in ac driven systems as we decrease the frequency,
because the distance between Floquet bands is proportional to $\omega$.
If the maximum width of an isolated Floquet band, is given by $\delta\epsilon\leq\omega$
($\epsilon_{\alpha}\in\left[-\omega/2,\omega/2\right]$). Then, the
Floquet bands at $\pm\omega$ close the gap when $\omega=\delta\epsilon/2$.
As a general rule, band inversions happen for every:
\begin{equation}
\omega_{n}=\frac{\delta\epsilon}{2n},\ n\in\mathbb{Z}^{+},\label{eq:GI-Freq}
\end{equation}
where $\mathbb{Z}^{+}$ denotes the set of positive integers. Therefore,
at $\omega_{n}$ the $\pm n\omega$ Floquet bands close the gap, switching
between an ordinary and a topological insulating phase. For example,
for a dimers chain with $b_{0}=0$ the maximum width coincides with
the undriven system band width, i.e., $\delta\epsilon=\delta E=2\tau\sqrt{1+\lambda^{2}+2\lambda}$.
Then, by means of Eq.\ref{eq:GI-Freq} it is possible to track the
bands inversions in terms of the undriven system\cite{Sup.inf.}.

\begin{figure}[h]
\includegraphics[scale=0.8]{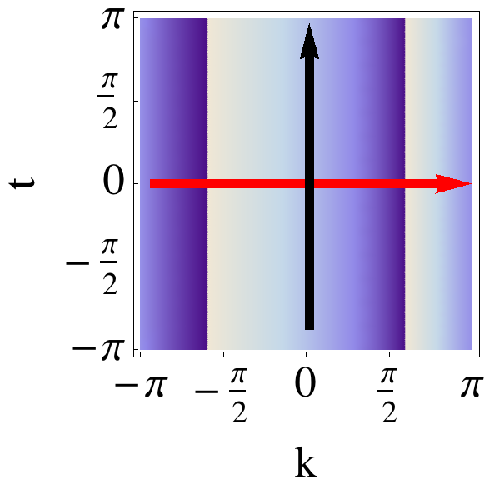}\includegraphics[scale=0.8]{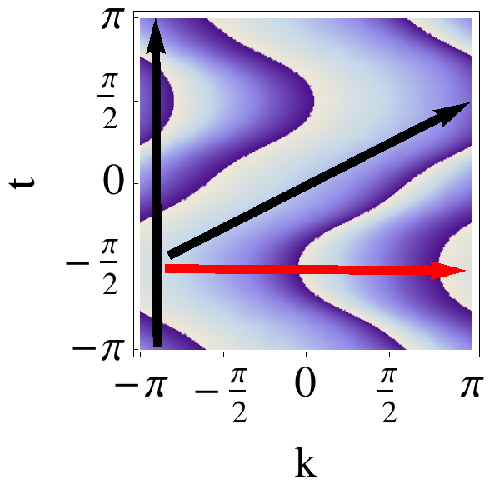}

\caption{\label{fig:Phase-divergences-1}Plot of $\phi\left(k,t\right)$ all
over the FFBZ for $A_{0}=0$ (left) and $A_{0}=2$ (right). Paths
parallel to $k$ cross two discontinuities meaning $\nu_{1}=\pm\pi$
(red arrow), on the other hand, paths parallel to $t$ have $\nu_{1}=0$
because they wind back and forth (black arrow).}
\end{figure}

For $\omega\ll\tau,\tau^{\prime}$, a large number of bands inversions
occur, being difficult to track all of them. In addition, the presence
of a D+1 base manifold becomes important. In that case, one can neglect
the time derivative in Eq.\ref{eq:Floquet-operator}, and diagonalize
the operator in $t$ domain. In that case, one obtains a 2 by 2 Floquet
operator $\mathcal{H}\left(k,t\right)\simeq H\left(k,t\right)_{\text{NN}}$
defined over the FFBZ, where $H\left(k,t\right)_{\text{NN}}$ extends
up to next nearest neighbors coupling in $\left(n,m\right)$ (it is
a good approximation for $A_{0}\leq1$). However, 2D Hamiltonians
in the BDI class are topologically trivial, and only the 1D topological
invariant $\nu_{1}$ is still non zero. It means that all changes
in the topological properties will be reflected in $\nu_{1}$. In
addition, for BDI Hamiltonians one can compute the winding number
graphically\cite{Pierre-Dimers}, in terms of the divergences of the
phase $\phi\left(k,t\right)=\arctan\left(g_{y}/g_{x}\right)$ in the
FFBZ (Fig.\ref{fig:Phase-divergences-1}), being $g_{x,y}$ the components
of the vector 
\begin{equation}
H\left(k,t\right)_{\text{NN}}=\tau\vec{g}\left(k,t\right)\cdot\vec{\sigma}.\label{eq:low-freq-H}
\end{equation}
 Note that in difference with Eq.\ref{eq:Zeroth-order-H}, $\vec{g}\left(k,t\right)$
now depends on $t$, and $\nu_{1}$ can be defined along the two inequivalent
axis of the torus
\[
\nu_{1}\left(\eta\right)=\frac{1}{2}\oint\frac{\partial}{\partial\mu}\phi\left(\mu,\eta\right)d\mu,\quad\mu,\eta=k,t.
\]
 
For the case of a finite chain, the existence of boundary states depends
on the loops along the $k$ axis, i.e., on $\nu_{1}\left(t\right)$\cite{Pierre-Dimers}.
In Fig.\ref{fig:Phase-divergences-1} $\nu_{1}\left(t\right)=\pi$
for all $t$, independently of the value of $E_{0}/\omega$ (trajectories
always cross two discontinuities). Thus, gap inversions are not relevant
for the existence of edge states in the low frequency regime, being
a feature of the transition from the FBZ to the FFBZ. This can be
seen numerically in the finite size system, in which zero energy modes
are present independently on the gap inversions\cite{Sup.inf.}.

\paragraph{Conclusions:}

We have derived a general approach to solve periodically driven D
dimensional lattices. It allows to obtain effective Hamiltonians for
the different driving regimes and a complete topological classification
in terms of AZ classes. We show that even for high frequency, the
underlying topology of the undriven system is present due to the time
periodicity. In addition, we show that at low frequency, the Floquet
operator is analog to the one of a static system in D+1 dimensions,
leading to interesting topological states of matter which otherwise
would be inaccessible. Finally, we also described the mechanism of
bands inversion in ac driven systems and its relation with the topology
of the system.

We have shown a large horizon of possibilities for periodically driven
systems, which in addition, can be studied experimentally e.g., by
measuring the electric polarizability or the appearance of boundary
states. The driving allows to simulate properties of undriven systems
in higher dimensions and the obtention of new topological phases due
to tunable hoppings\cite{NNN}. One could also think in more exotic
types of zero energy modes in the low frequency regime, as for example
those in the boundary between a driven and undriven materials. Also,
further studies in the case of bichromatic ac fields could be interesting.
Finally, the results in the high frequency regime can be used for
non-adiabatic topological quantum computation.

We thank P. Delplace, S. Kohler, P.C.E. Stamp, and M. Büttiker for
useful discussions. A.G.L. acknowledges the JAE program (MICINN) and
we both acknowledge Grant No. MAT2011-24331 and the ITN Grant No.
234970 (EU).

\section{Suplementary Information}

\section*{S.1: Relation between undriven and ac driven Hamiltonians}

For a periodic lattice, driven by an ac electric field under the dipolar
approximation ($\mathbf{A}\left(\mathbf{x},t\right)\simeq\mathbf{A}\left(t\right)$),
the relation between the static Hamiltonian and the ac driven Hamiltonian
can be obtained by noticing that: 
\begin{eqnarray}
H & = & \frac{\mathbf{p}^{2}}{2m}+V\left(\mathbf{x}\right),\label{eq:relations-hamiltonians-1}\\
H_{\mathbf{k}} & \equiv & e^{-i\mathbf{k}\cdot\mathbf{x}}He^{i\mathbf{k}\cdot\mathbf{x}}=\frac{\left(\mathbf{p}+\mathbf{k}\right)^{2}}{2m}+V\left(\mathbf{x}\right),\nonumber \\
H\left(t\right) & = & \frac{\left(\mathbf{p}+\mathbf{A}\left(t\right)\right)^{2}}{2m}+V\left(\mathbf{x}\right),\nonumber \\
H_{\mathbf{k}}\left(t\right) & \equiv & e^{-i\mathbf{k}\cdot\mathbf{x}}H\left(t\right)e^{i\mathbf{k}\cdot\mathbf{x}}=\frac{\left(\mathbf{p}+\mathbf{k}+\mathbf{A}\left(t\right)\right)^{2}}{2m}+V\left(\mathbf{x}\right)\nonumber \\
 & = & e^{-i\left(\mathbf{k}+\mathbf{A}\left(t\right)\right)\cdot\mathbf{x}}He^{i\left(\mathbf{k}+\mathbf{A}\left(t\right)\right)\cdot\mathbf{x}}.\nonumber 
\end{eqnarray}
It relies in the minimal coupling $\mathbf{k}\rightarrow\mathbf{K}\left(t\right)=\mathbf{k}+\mathbf{A}\left(t\right)$.
Explicitly for the case of a tight binding Hamiltonian, one should
notice that the hoppings can be defined as the Fourier transform of
the energy:
\[
\tau_{j,l}\equiv N^{-D}\sum_{\mathbf{k}}E_{\mathbf{k}}e^{-i\mathbf{k}\cdot\left(\mathbf{R}_{j}-\mathbf{R}_{l}\right)},
\]
which are $\mathbf{k}$ independent. Therefore, the inverse Fourier
transform of $\tau_{j,l}$ can encode the time dependence in the minimal
coupling as:
\[
H_{\mathbf{k}}\left(t\right)=N^{-D}\sum_{j,l}\tau_{j,l}e^{i\mathbf{K}\left(t\right)\cdot\left(\mathbf{R}_{j}-\mathbf{R}_{l}\right)}c\left(t\right)_{\alpha,\mathbf{k}}^{\dagger}c\left(t\right)_{\alpha,\mathbf{k}},
\]
where the time dependent hoppings $\tau_{j,l}\left(t\right)=\tau_{j,l}e^{i\mathbf{A}\left(t\right)\cdot\left(\mathbf{R}_{j}-\mathbf{R}_{l}\right)}$
are obtained from the time independent ones. This result relies on
the same principle as the Bloch equation for a particle in a dc electric
field\cite{Bloch-eq}.

\section*{S.2: 1D system in low frequency regime}

Let us consider as an example, a 1D chain driven by an ac electric
field, where $\omega\ll\tau_{i,j}$, being $\tau_{i,j}$ the hopping
between sites, and $\omega$ the frequency of the ac field. In this
case, one can neglect the time derivative term of the Floquet operator,
or equivalently the effective static electric field, as we described
in the main text. Then, the translational symmetry along the energy
axis $\mathcal{E}$ is recovered, and one can write the Floquet operator
in $t$ domain, where it is diagonal:
\begin{eqnarray}
\mathcal{H}\left(\mathbf{k},t\right) & \simeq & M^{-1}\sum_{\alpha,\mathbf{k}}\sum_{m,n}\tau_{n,m}c\left(t\right)_{\alpha,\mathbf{k}}c\left(t\right)_{\alpha,\mathbf{k}}^{\dagger}e^{i\omega t\left(n-m\right)},\label{eq:Floquet_op-low}
\end{eqnarray}
being $M$ a normalization factor. Thus, because the Floquet operator
(Eq.\ref{eq:Floquet_op-low}) depends on the compact parameter space
$\left(k,t\right)$, we can classify the topological properties of
$\mathcal{H}\left(k,t\right)$ according to the AZ classification
of time independent D+1 dimensional systems\cite{topological-clasification-1}.
It classifies the mappings from the torus to the family of Floquet
operators $\mathcal{H}\left(k,t\right)$. For example, the case of
a 2 band model with Floquet operator:
\[
\mathcal{H}\left(k,t\right)=\vec{h}\left(k,t\right)\cdot\vec{\sigma},
\]
possess a first Chern number given by:
\begin{eqnarray*}
c_{1} & = & \int_{FBZ}\int_{-\frac{\pi}{\omega}}^{\frac{\pi}{\omega}}\hat{h}\left(k,t\right)\cdot\left(\frac{\partial}{\partial k}\hat{h}\left(k,t\right)\times\frac{\partial}{\partial t}\hat{h}\left(k,t\right)\right)dtdk\\
 & = & \int_{FFBZ}\hat{h}\left(k,t\right)\cdot\left(\frac{\partial}{\partial k}\hat{h}\left(k,t\right)\times\frac{\partial}{\partial t}\hat{h}\left(k,t\right)\right)dtdk,
\end{eqnarray*}
being the First Floquet Brillouin Zone (FFBZ) homeomorphic to a torus
$\mathbb{T}^{1+1}$, and $\hat{h}=\vec{h}/|\vec{h}|$. Note that for
a one dimensional time independent system, $c_{1}$ would always vanish
because the FBZ is given by $\mathbb{T}^{1}$. However, the increase
of the parameter space dimension, due to the time periodicity, allows
for higher order topological invariants.

In the particular case of a dimers chain, discussed in the main text,
$c_{1}=0$. The reason is that the Floquet operator belongs to the
BDI class in 2D. Therefore, the winding number $\nu_{1}$ is the topological
invariant that differentiates our system from an ordinary insulator.

\section*{S.3 Dimers chain}

\section*{S.3.1: Hamiltonians and topological invariants}

The undriven tight binding model for nearest neighbors is given by:
\begin{eqnarray}
H_{k} & = & \left(\begin{array}{cc}
0 & \rho\left(k\right)\\
\rho\left(k\right)^{*} & 0
\end{array}\right),\label{eq:Undriven-Dimers}\\
\rho\left(k\right) & \equiv & \tau^{\prime}e^{-ikb_{0}}+\tau e^{ik\left(a_{0}-b_{0}\right)},\nonumber 
\end{eqnarray}
\[
E_{\pm}=\pm\tau\sqrt{\lambda^{2}+1+2\lambda\cos\left(ka_{0}\right)},
\]
being $\lambda\equiv\tau^{\prime}/\tau$ (see Fig.2 in the main text).
Importantly, the dispersion relation does not depend on $b_{0}$,
and for the condition $b_{0}=a_{0}/2$, and $\tau=\tau^{\prime}$
in $H_{k}$ one recovers the energy spectrum of the linear chain.

\begin{widetext}

\begin{figure}[h]
\includegraphics[scale=0.6]{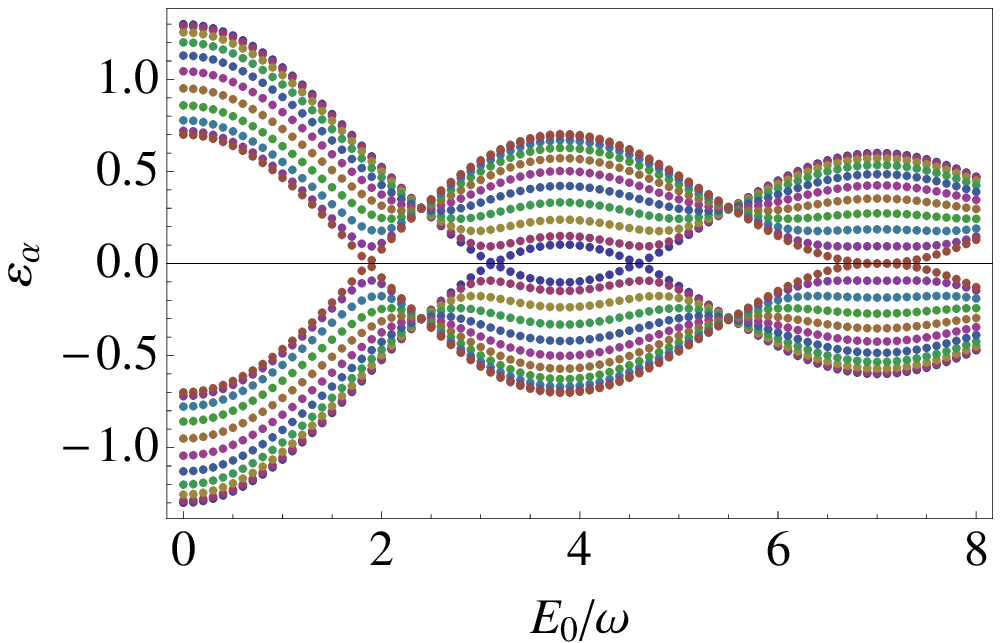}\includegraphics[scale=0.6]{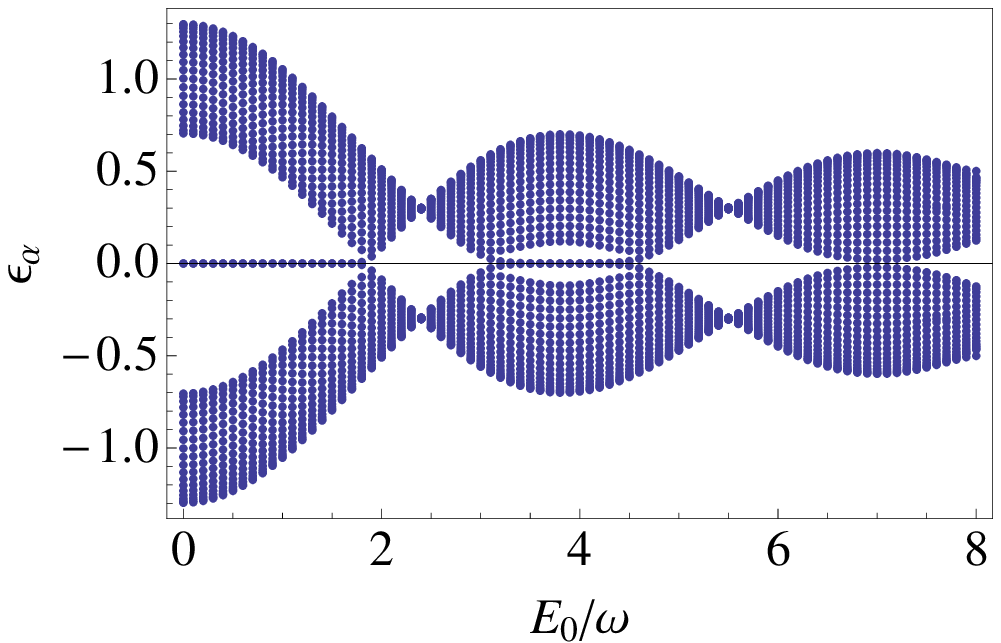}

\caption{\label{fig:Quasi-energies-vs-Intensity1-1}Quasi-energy spectrum vs
$E_{0}/\omega$ for $\lambda=0.3$, and $b_{0}=0$, considering Eq.\ref{eq:Quasienergy-Dimers1-1}
for the model with periodic boundary conditions (left), and the numerical
solution for the finite system with 20 dimers (right). For the finite
dimers chain we have considered $n,m=150$ sidebands and $\omega=10$.
Note the existence of gapless modes for a certain range of $E_{0}/\omega$.
All parameters in units of $\tau=1$.}
\end{figure}

\end{widetext}

\begin{widetext}

\begin{figure}[h]
\includegraphics[scale=0.65]{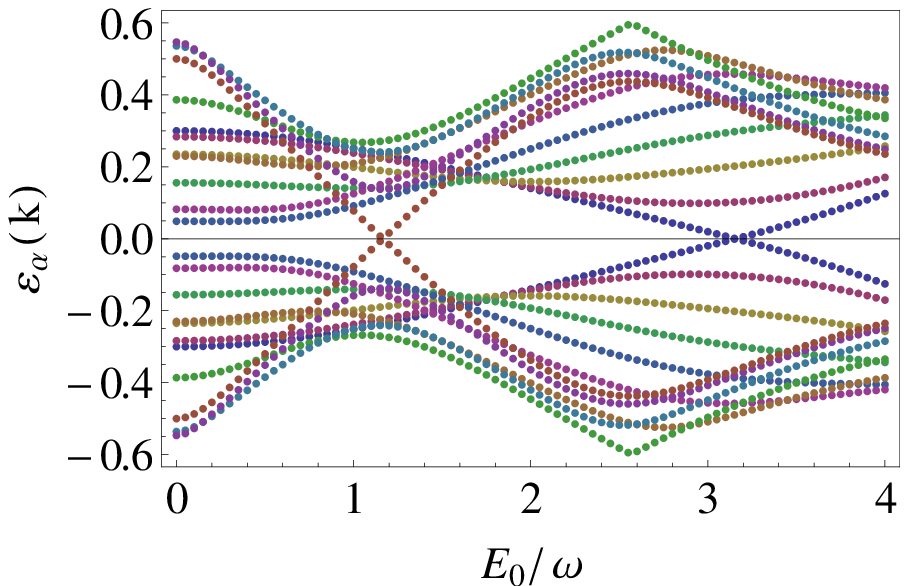}\includegraphics[scale=0.65]{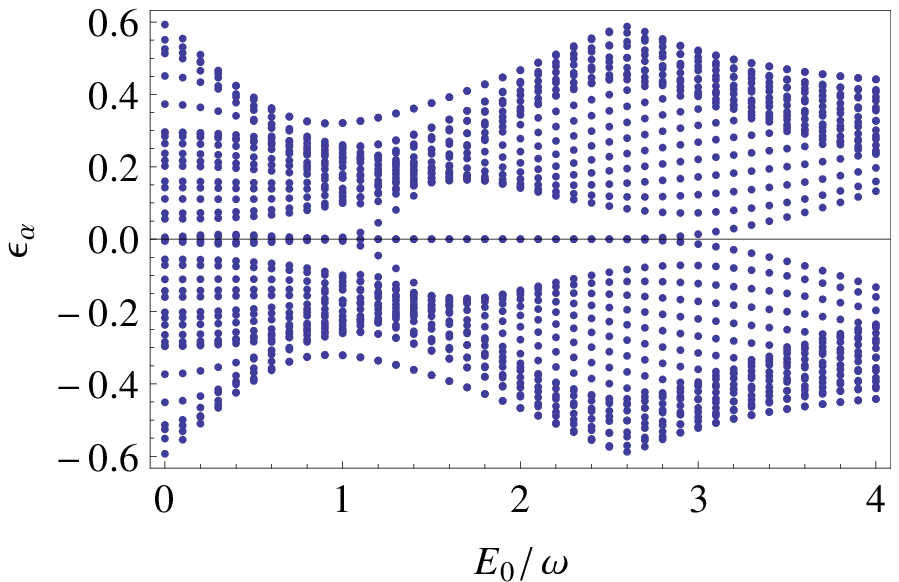}

\caption{\label{fig:Quasi-energy-vs-E-Resonance}Quasi-energies vs $E_{0}/\omega$
for $\omega\simeq\tau,\tau^{\prime}$. We compare the model with periodic
boundary conditions (left) and the finite one (right). Both models
only differ in the finite size effects. The periodic boundary model
has considered $m,n=15$ and up to fifteen order coupling, while the
finite size model has considered $m,n=85$. Parameters $\omega=1.2$,
$\lambda=0.5$, and $b_{0}=0$ (all parameters in units of $\tau=1$).}
\end{figure}

\end{widetext}

We consider the ac vector potential $A\left(t\right)=A_{0}\sin\left(\omega t\right)$,
being $A_{0}\equiv qE_{0}/\omega$. By means of the minimal coupling
we arrive at the time dependent Hamiltonian:
\begin{eqnarray*}
H_{K\left(t\right)} & = & \tau\left(\begin{array}{cc}
0 & \rho\left(k,t\right)\\
\rho\left(k,t\right)^{*} & 0
\end{array}\right),\\
\rho\left(k,t\right) & \equiv & \lambda e^{-i\left(k+A_{0}\sin\left(\omega t\right)\right)b_{0}}+e^{i\left(k+A_{0}\sin\left(\omega t\right)\right)\left(a_{0}-b_{0}\right)}.
\end{eqnarray*}
In order to calculate $\tilde{\tau}_{n,m}$ (Eq.7 in the main text),
we use the identity $J_{n}\left(x\right)=\frac{1}{T}\int_{0}^{T}e^{ix\sin\left(\omega t\right)-ip\omega t}dt$,
leading to:
\begin{eqnarray}
\tilde{\tau}_{n,m} & = & \tau\left(\begin{array}{cc}
0 & \rho_{F}\left(k\right)\\
\tilde{\rho}_{F}\left(k\right) & 0
\end{array}\right),\label{eq:hopping-dimers}
\end{eqnarray}
\begin{eqnarray}
\rho_{F}\left(k\right) & \equiv & \lambda e^{-ikb_{0}}J_{n-m}\left(A_{0}b_{0}\right)\label{eq:rho-effectives}\\
 &  & +e^{ik\left(a_{0}-b_{0}\right)}J_{m-n}\left(A_{0}\left(a_{0}-b_{0}\right)\right),\nonumber \\
\tilde{\rho}_{F}\left(k\right) & \equiv & \lambda e^{ikb_{0}}J_{m-n}\left(A_{0}b_{0}\right)\nonumber \\
 &  & +e^{-ik\left(a_{0}-b_{0}\right)}J_{n-m}\left(A_{0}\left(a_{0}-b_{0}\right)\right).\nonumber 
\end{eqnarray}
Finally, one can obtain the matrix elements of the Floquet operator
as:
\begin{equation}
\mathcal{H}_{k}^{\left(n,m\right)}=\tilde{\tau}_{n,m}-n\omega\delta_{n,m},\label{eq:FloquetEq-FourierSpace}
\end{equation}
where we have included the time derivative operator in Fourier space.
Note that this is an infinite matrix because $\left(n,m\right)\in\mathbb{Z}$.
In the high frequency regime ($\omega\gg\tau,\tau^{\prime}$), the
second term in the right hand side of Eq.\ref{eq:FloquetEq-FourierSpace}
dominates, and the matrix is approximately block diagonal. Thus, we
select the Floquet band $m=n=0$ for simplicity, being the effective
Hamiltonian given by the 2 by 2 matrix:
\begin{eqnarray}
\mathcal{H}_{k}^{\left(0\right)} & = & \tau\left(\begin{array}{cc}
0 & \rho_{F}^{\left(0\right)}\\
\left(\rho_{F}^{\left(0\right)}\right)^{*} & 0
\end{array}\right),\nonumber \\
\rho_{F}^{\left(0\right)} & \equiv & \lambda J_{0}\left(A_{0}b_{0}\right)+e^{ika_{0}}J_{0}\left(A_{0}\left(a_{0}-b_{0}\right)\right).\label{eq:rho-effectives-zero}
\end{eqnarray}
Note that we have considered $\rho_{F}^{\left(0\right)}$ in a different
basis than in Eq.\ref{eq:rho-effectives}. Both basis differ in a
phase factor $e^{\pm ikb_{0}}$. The reason is that in order to properly
obtain the topological properties we must consider the closed lifts
basis\cite{Montambaux-tight_binding,Mostafazadeh}. The quasi-energy
spectrum at high frequency is ($q=-e=-1$):\begin{widetext}
\begin{eqnarray}
\epsilon_{\pm,k}^{0} & = & \pm\tau\sqrt{\lambda^{2}J_{0}^{2}\left(x\right)+J_{0}^{2}\left(y\right)+2\lambda\cos\left(ka_{0}\right)J_{0}\left(x\right)J_{0}\left(y\right)},\label{eq:Quasienergy-Dimers1-1}
\end{eqnarray}

\end{widetext}where $x\equiv A_{0}b_{0}$, and $y\equiv A_{0}\left(a_{0}-b_{0}\right)$.
Fig.\ref{fig:Quasi-energies-vs-Intensity1-1} shows a comparison between
the numerical calculation of a finite size system in high frequency
regime (right), and the quasi-energies obtained in the high frequency
regime for the model with periodic boundary conditions, Eq.\ref{eq:Quasienergy-Dimers1-1}
(left).Out of the high frequency regime the Floquet bands
couple, and we must solve the full Floquet equation. Fig.\ref{fig:Quasi-energy-vs-E-Resonance}
shows a comparison between the model with periodic boundary conditions
(left) and the finite one (right), for frequency $\omega\simeq\tau,\tau^{\prime}$.
In this regime, the Floquet bands couple and the quasi-energies obtained
in Eq.\ref{eq:Quasienergy-Dimers1-1} are no longer valid. Then, we
numerically diagonalize Eq.\ref{eq:FloquetEq-FourierSpace} for $n,m=15$
sidebands, and compare with the calculation for the finite system,
which requires a larger number of Floquet bands in order to reach
convergence ($n,m=85$).The full Floquet operator in time domain is given by
\begin{equation}
\left(H\left(k,t\right)-i\partial_{t}\right)|u_{\alpha,k}\rangle=\epsilon_{\alpha,k}|u_{\alpha,k}\rangle.\label{eq:Diff-Eq-Floquet}
\end{equation}
For the calculation of $\mathcal{H}\left(k,t\right)$, the coupling
between different Floquet bands depends on the amplitude of the vector
potential. If we assume $A_{0}\leq1$, we can neglect the contributions
from $J_{p>2}\left(A_{0}\right)$, leading to a Hamiltonian up to
next nearest neighbors:
\begin{eqnarray}
H\left(k,t\right)_{\text{NN}} & = & \tau\left(\begin{array}{cc}
0 & \rho_{F}\left(k,t\right)\\
\rho_{F}\left(k,t\right)^{*} & 0
\end{array}\right),\label{eq:Floquet-op-Low-freq}\\
\rho_{F}\left(k,t\right) & \equiv & \lambda J_{0}\left(x\right)+2i\lambda J_{1}\left(x\right)\sin\left(\omega t\right)\nonumber \\
 &  & +2\lambda J_{2}\left(x\right)\cos\left(2\omega t\right)+e^{ika_{0}}J_{0}\left(y\right)\nonumber \\
 &  & -2e^{ika_{0}}\left(iJ_{1}\left(y\right)\sin\left(\omega t\right)-J_{2}\left(y\right)\cos\left(2\omega t\right)\right),\nonumber 
\end{eqnarray}
which, in the limit $A_{0}\rightarrow0$ becomes the energy dispersion
of the undriven system.

Eq.\ref{eq:Diff-Eq-Floquet} is not exactly solvable for the time
dependent Hamiltonian $H\left(k,t\right)_{\text{NN}}$ (Eq.\ref{eq:Floquet-op-Low-freq}).
However, for frequency values $\omega\ll\tau,\tau^{\prime}$ we can
neglect the time derivative, such that $\mathcal{H}\left(k,t\right)\simeq H\left(k,t\right)_{\text{NN}}$.
Then, Eq.\ref{eq:Diff-Eq-Floquet} describes an eigenvalue equation,
and we can classify the mappings $\left(k,t\right)\rightarrow\mathcal{H}\left(k,t\right)$,
from the FFBZ to the set of Floquet operators: 
\begin{equation}
\mathcal{H}\left(k,t\right)=\tau\vec{g}\left(k,t\right)\cdot\vec{\sigma}.\label{eq:Adiabatic-HF}
\end{equation}
The calculation of the first Chern number using Eq.\ref{eq:Adiabatic-HF}
gives $c_{1}=0$, because the Hamiltonian belongs to the BDI class.
However, we can calculate the winding number $\nu_{1}$ along the
two inequivalent directions of the 2-torus $k$ and $t$ as: $\nu_{1}\left(\eta\right)=\frac{1}{2}\oint\frac{\partial}{\partial\mu}\phi\left(\mu,\eta\right)d\mu$,
where $\mu,\nu=k,t$, $\phi\left(k,t\right)\equiv\arctan\left(\frac{g_{y}}{g_{x}}\right)$,
and $g_{x,y}$ are the components of the vector $\vec{g}\left(k,t\right)$
in Eq.\ref{eq:Adiabatic-HF}. As we discussed in the main text, in
this case $\nu_{1}\left(t\right)=\pi$ for all $t$, and $\nu_{1}\left(k\right)=0$
for all $k$.

\begin{widetext}

\begin{figure}[H]
\includegraphics[scale=0.65]{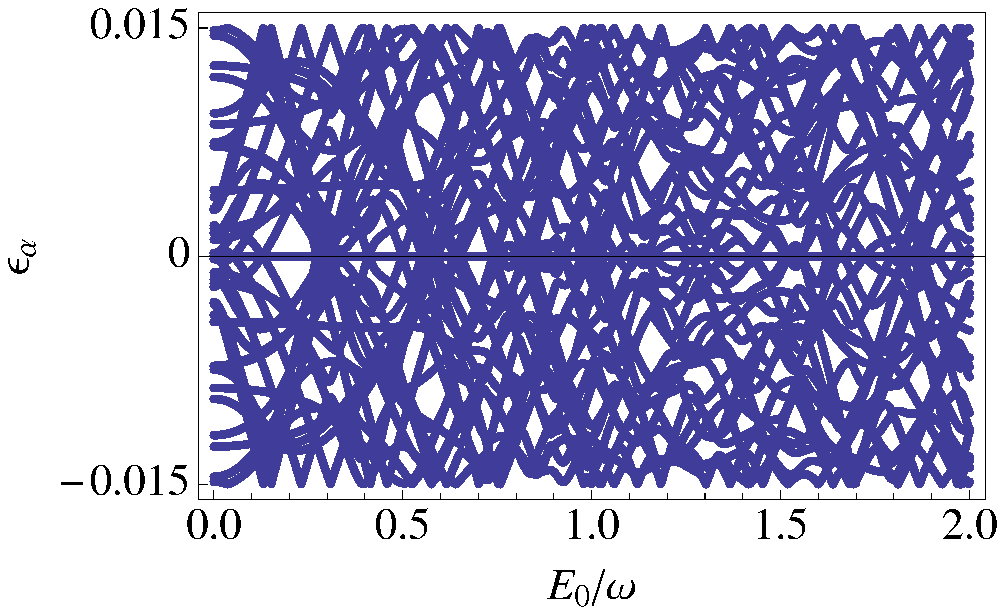}\includegraphics[scale=0.65]{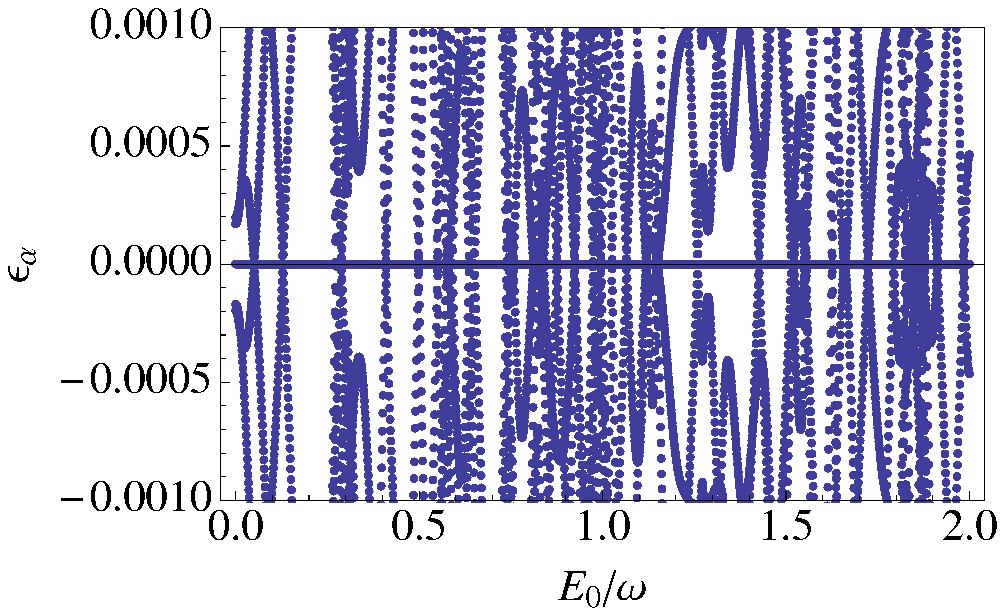}

\caption{\label{fig:low-frequency}(Left) Quasi-energy spectrum within $\left(-\omega/2,\omega/2\right)$
for a finite dimers chain in the low frequency regime ($\omega\ll\tau,\tau^{\prime}$).
Right plot shows a zoom of the zero energy mode. The plots show a
large number of crossings between the quasi-energies. However, the
existence of zero energy modes is not affected by these crossings,
and they exists for all values of $E_{0}/\omega$. We have considered
a finite chain of 20 dimers, $n,m=105$, $\omega=3.10^{-2}$, and
$\lambda=0.3$ in units of $\tau$.}
\end{figure}

\end{widetext}

In the present example of a dimer chain, the existence of topologically
protected states at the boundary of the chain is related with the
loops along $k$, i.e., with the winding number $\nu_{1}\left(t\right)$.
Hence, $\nu_{1}\left(t\right)=\pi$ for all $t$ means the existence
of boundary states for all $t$, independently of the value of $A_{0}$,
as far as we are in the low frequency regime (see Fig.\ref{fig:low-frequency}).

\section*{S.3.2: Bands Inversions}

\begin{widetext}

\begin{figure}[H]
\includegraphics[scale=0.6]{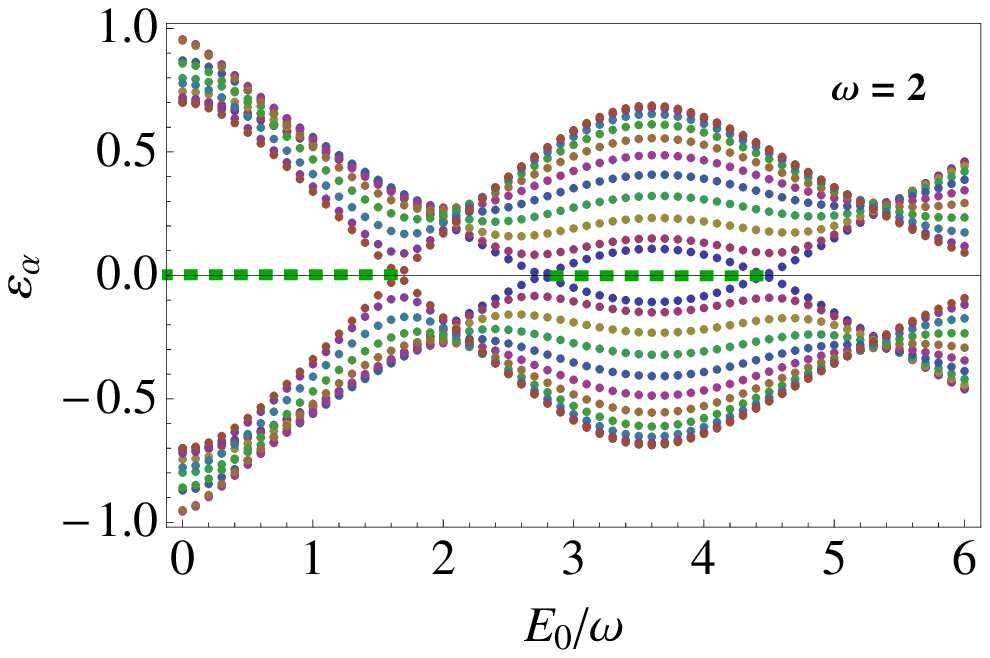}\includegraphics[scale=0.6]{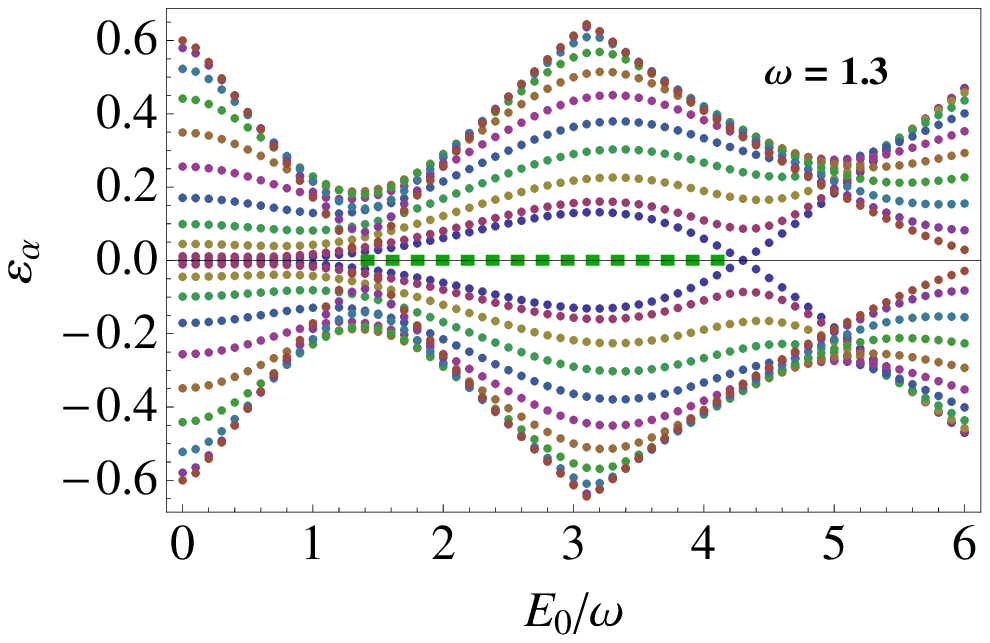}\includegraphics[scale=0.6]{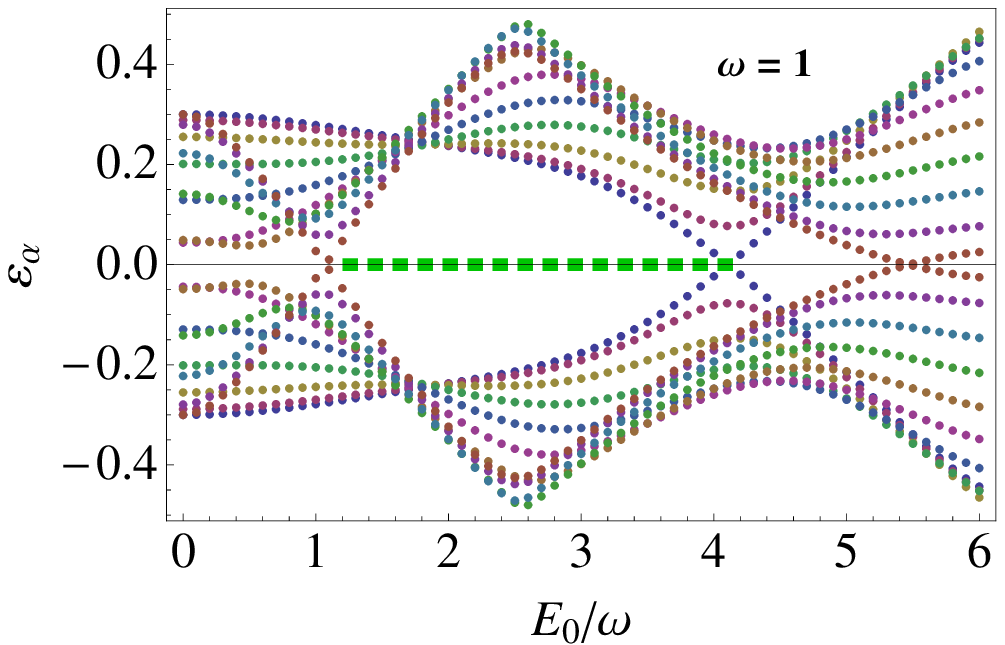}

\caption{\label{fig:topology-resonance-1}Quasi-energies vs $A_{0}=E_{0}/\omega$
for different frequencies considering periodic boundary conditions.
In dotted green color we plot the boundary states, which has been
obtained using a finite tight binding model. Left figure shows the
quasi-energies for $\omega=2$, where the bands are not inverted.
The center figure shows the quasi-energies for the critical value
of $\omega_{1}=1.3$ at which the Floquet bands that belong to $\pm\omega$
close the gap at $A_{0}=0$. The right figure considers a lower frequency
($\omega=1$) where the gap is reopened. Importantly, the comparison
between left and right figures shows that the order in which the boundary
states appear as a function of $A_{0}$ is inverted.}
\end{figure}

\end{widetext}

Bands inversions in periodically driven lattices are produced due
to the coupling between different Floquet bands as we decrease the
frequency. They change the topological properties of the system by
opening and closing the gap. Fig.\ref{fig:topology-resonance-1} shows
the band structure of the dimers chain for different values of $\omega$.
The exact crossings between conduction and valence band in this model
are critical points for a topological phase transition, as the appearance/disappearance
of boundary states demonstrate. While the critical points for $\omega=2$
(Fig.\ref{fig:topology-resonance-1} left) at $E_{0}/\omega\simeq1.5\text{, and }4.5$
are almost fixed as we decrease the frequency, the critical point
at $E_{0}/\omega\sim2.8$ shifts to the left as the frequency is lowered
(Fig.\ref{fig:topology-resonance-1} center and right). It reaches
the origin at $\omega_{1}=1.3$ and vanishes modifying the phase diagram.
Importantly, the shift of the critical points depends on the band
structure of the system under consideration (the critical points can
disappear at the origin, or new ones can show up). However, the gap
always closes for $\omega_{n}$, independently of the critical points
dynamics (being $\omega_{n}\equiv\delta\epsilon/n\omega$, $n\in\mathbb{Z}^{+}$,
and $\delta\epsilon$ the maximum thickness of the Floquet band).

\end{document}